# Phonon scattering in ortho-para hydrogen solid solutions (role of configurational relaxation)

B. Ya. Gorodilov*

*B. Verkin Institute for Low Temperature Physics and Engineering, National Academy of Sciences of Ukraine, pr. Lenina, 47, 61103 Kharkov, Ukraine*

The experimental data on the thermal conductivity of ortho-parahydrogen solutions are analyzed on the basis of a relaxation-time model taking account of configurational relaxation of the ortho subsystem. The influence of configurational relaxation on the thermal conductivity is analyzed using resonance scattering of phonons by pair clusters of orthomolecules taking account of their rotational spectrum.

## 1. INTRODUCTION

Solid parahydrogen is a convenient object for investigating heat transfer, since it is quite easy to obtain a crystal with the minimal number of defects and phonon mean-free path of the order of the dimensions of the crystal.[1] At the same time ortho-para hydrogen solutions make it possible to investigate phonon scattering by the rotational motion of the molecules without additional scattering effects due to the difference in the masses and interaction potentials. Orthomolecules in parahydrogen are impurities which differ from the matrix only by the presence of a nonzero rotational angular momentum.

Investigations of the thermal conductivity of parahydrogen have revealed a number of peculiarities in heat transfer processes in this crystal:

—the intensity of normal scattering processes in high-quality crystals is comparable to that of other scattering processes, so that they cannot be neglected when experimental results are analyzed;[2]

—the thermal conductivity exhibits anisotropy due to *hcp*-crystal orientation;[3]

—pure parahydrogen and a crystal with an impurity behave differently under thermal deformations: deformation of a pure crystal results in additional thermal resistivity with inverse-square dependence on the temperature $T$, whereas in an impure crystal the additional thermal resistivity is proportional to $T^3$.[4]

The first measurements of the thermal conductivity of solid hydrogen, which were performed in Ref. 5, revealed a strong dependence of the thermal conductivity of solid ortho-parahydrogen solutions on the content of the ortho modification in the sample. Subsequent measurements[6] showed that the additional thermal resistivity due to the orthomolecules exhibits a temperature dependence close to $T^{-3}$ and a quadratic concentration dependence. At the same time a calculation of the influence of the ortho impurity on the thermal conductivity performed in Ref. 7 gave a temperature dependence proportional to $T^2$ for the effect. Taking account of phonon scattering by a pair of orthomolecules[8] made it possible to describe the experiment of Ref. 6 but this required substantial scaling of the computed effect. Elaboration of the theory of Ref. 7 has shown that at temperatures below 1 K the thermal resistivity due to orthomolecules is proportional $T^7$.[9,10] Subsequent experimental studies of ortho-parahydrogen solutions at temperatures below 4 K revealed relaxation of the thermal conductivity; this was explained by a change in the spatial arrangement of the orthomolecules in the crystal (configurational relaxation).[11,12] Configurational relaxation occurs in ortho-para hydrogen solutions because it is energetically more favorable for orthomolecules to occupy neighboring lattice sites, the energy gain in this case being about 4 K (Fig. 1, Ref. 13). The observed effect is quite large, and at 1.5 K for an approximately 2% concentration of orthomolecules it exceeded 40% of the thermal conductivity. This provided an impetus for revising the results obtained in Ref. 8. The authors of Ref. 14 performed a systematic microscopic analysis of phonon scattering by an impurity center consisting of a pair of orthomolecules but the computed effect was found to be negligible, much weaker than scattering by single orthomolecules (singles).

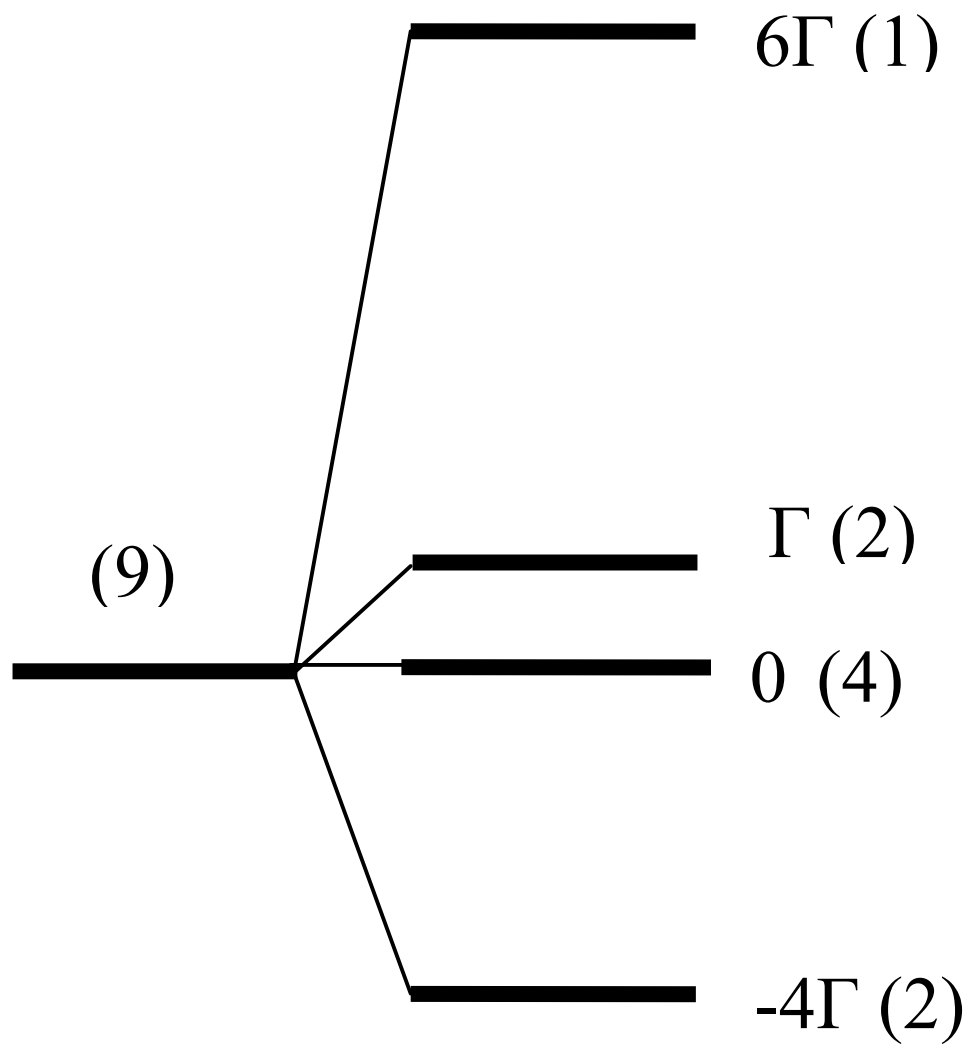

FIG. 1. Structure of the rotational levels of a pair of orthomolecules [Γ= 0.83 K (Ref. 17)].[13]

It would appear that other explanations for the relaxation of thermal conductivity should be sought. This would be true if it were not for the following circumstance. The relaxation of thermal conductivity observed in Refs. 11 and 12 differed in sign. In Ref. 11 an increase of thermal conductivity with time was observed at 0.2 K, whereas in Ref. 12 the thermal conductivity above 1.5 K was found to decrease with time. Subsequent investigations[15] showed that the relaxation of the thermal conductivity changes sign at a temperature close to 0.3 K (Fig. 2, Ref. 16), and the thermal conductivity relaxation times are the same as the relaxation times obtained for the ortho subsystem by NMR measurements performed in the same experiment. These investigations indicated unequivocally that the observed effect depends on the phonon frequencies corresponding to the experimental temperature. The phonon energies which determine the behavior of the thermal conductivity at each temperature correlate with the rotational spectrum of orthomolecules. The crystal-field splitting of the rotational levels of the singles ($J$=1) is close to 0.2 K,[17] and the smallest energy gap in the spectrum of an isolated pair of orthomolecules is about 1 K.[13] The centers that efficiently scatter phonons change as temperature decreases. Scattering by pair clusters, which determines the behavior of a crystal at temperatures close to and above 1 K, becomes weaker and essentially vanishes at 0.3 K. In summary, at present no theory describes the behavior of the thermal conductivity of ortho-para solutions of hydrogen.

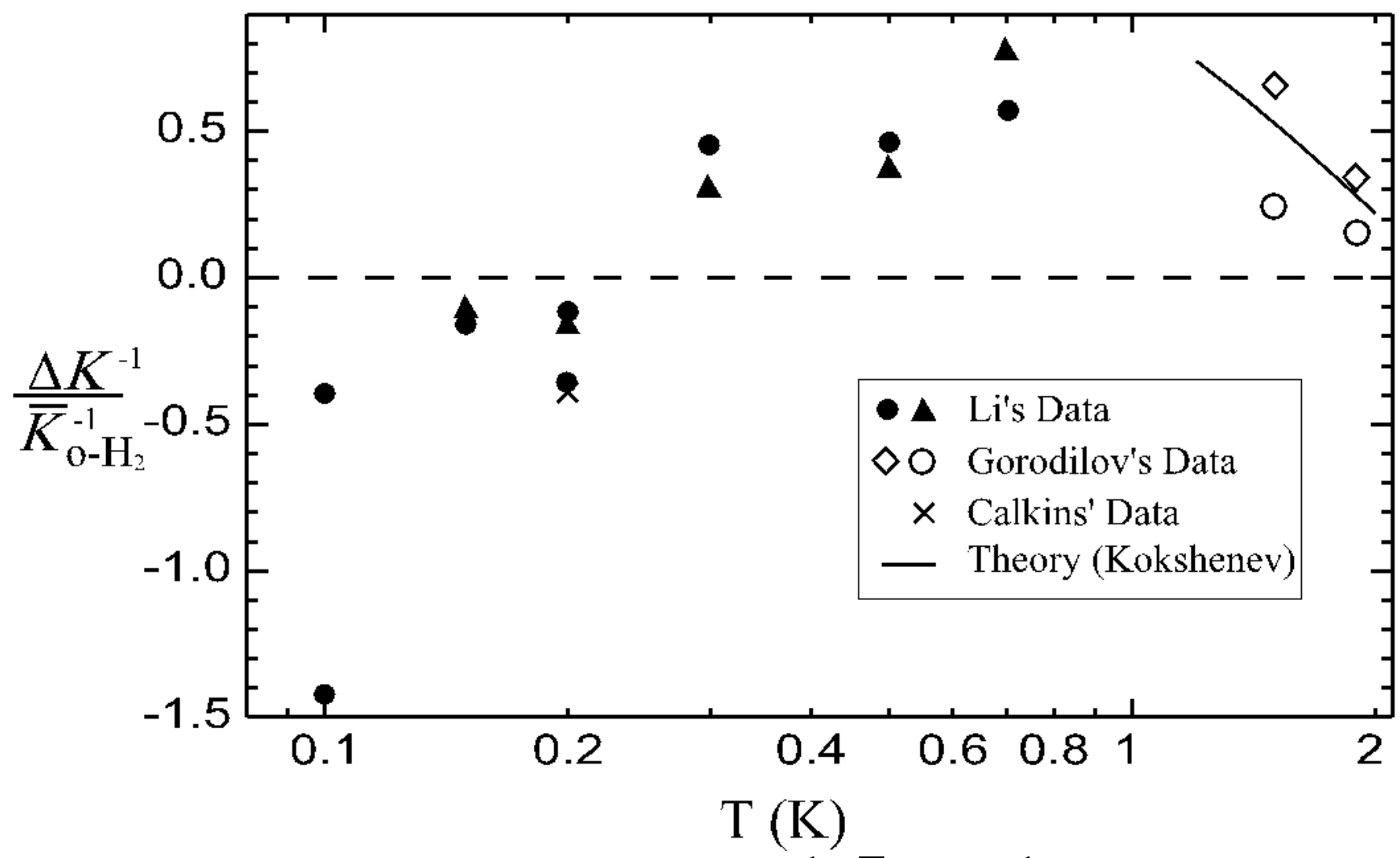

FIG. 2. The normalized change $\Delta K^{-1}/\ \bar{K}_{(J=1)H2}^{-1}$ versus $T$ for two series of experiments where $\bar{K}_{(J=1)H2}^{-1}$ is the average resistivity of the $(J = 1)H_2$. Close symbols $1.5\% < X < 3.4\%$ and $0.9\% < X < 1.1\%$.[15] Open symbols for $X = 1$ and 2%.[12] Cross $1.0\% < X < 2.5\%$.[11] The solid line is the prediction by Kokshenev[8] for $X = 1\%$.[16]

In the present paper an empirical model of thermal conductivity is proposed. It is shown that the relaxation effect can be described using the resonance scattering of phonons by pairs of orthomolecules and the change in the number of such molecules as a result of configurational relaxation.

## 2. MODEL OF THERMAL CONDUCTIVITY

Since normal scattering processes in parahydrogen ($N$-processes) are comparable to resistive processes, neither the Ziman limit, where the normal processes are much stronger than the resistive processes, nor the classical model, where the $N$-processes can be neglected, can be used to describe the thermal conductivity. The thermal conductivity of crystals with quasiequilibrium[12] and random distributions of the orthomolecules has been analyzed on the basis of the complete Callaway integral.[18] Figure 3 shows the experimental data of Ref. 12 on the temperature dependence of the thermal conductivity $K$ of ortho-para solutions of hydrogen for various concentrations $x$ (filled symbols); the open symbols represent the computed thermal conductivity of the unrelaxed crystal with a random distribution of the orthomolecules.[19]

The total thermal conductivity of a crystal can be represented as a sum of two integrals:

$$K(T) = GT^3 \left( \int_0^{\Theta/T} \tau_C f(x)dx + \frac{\int_0^{\Theta/T} \tau_C/\tau_N \; f(x)dx}{\int_0^{\Theta/T} \tau_C/\tau_N\tau_R \; f(x)dx} \right)$$

where

$$G = k^4/(2\pi^2 s\hbar^3); \qquad f(x) = x^4 e^x/(e^x - 1)^2;$$

$\Theta$ is the Debye temperature; $k$ is Boltzmann constant; $s$ is the sound speed; integration over the phonon frequency $\omega$ is replaced by integration over the dimensionless parameter $x = \hbar\omega/kT$; $\tau_R$ is the relaxation time for resistive scattering processes; $\tau_N$ is the relaxation time for $N$ processes; and, $\tau_C^{-1} = \tau_R^{-1} + \tau_N^{-1}$ is the combined relaxation time.

The intensity of normal phonon-phonon scattering processes has been determined in

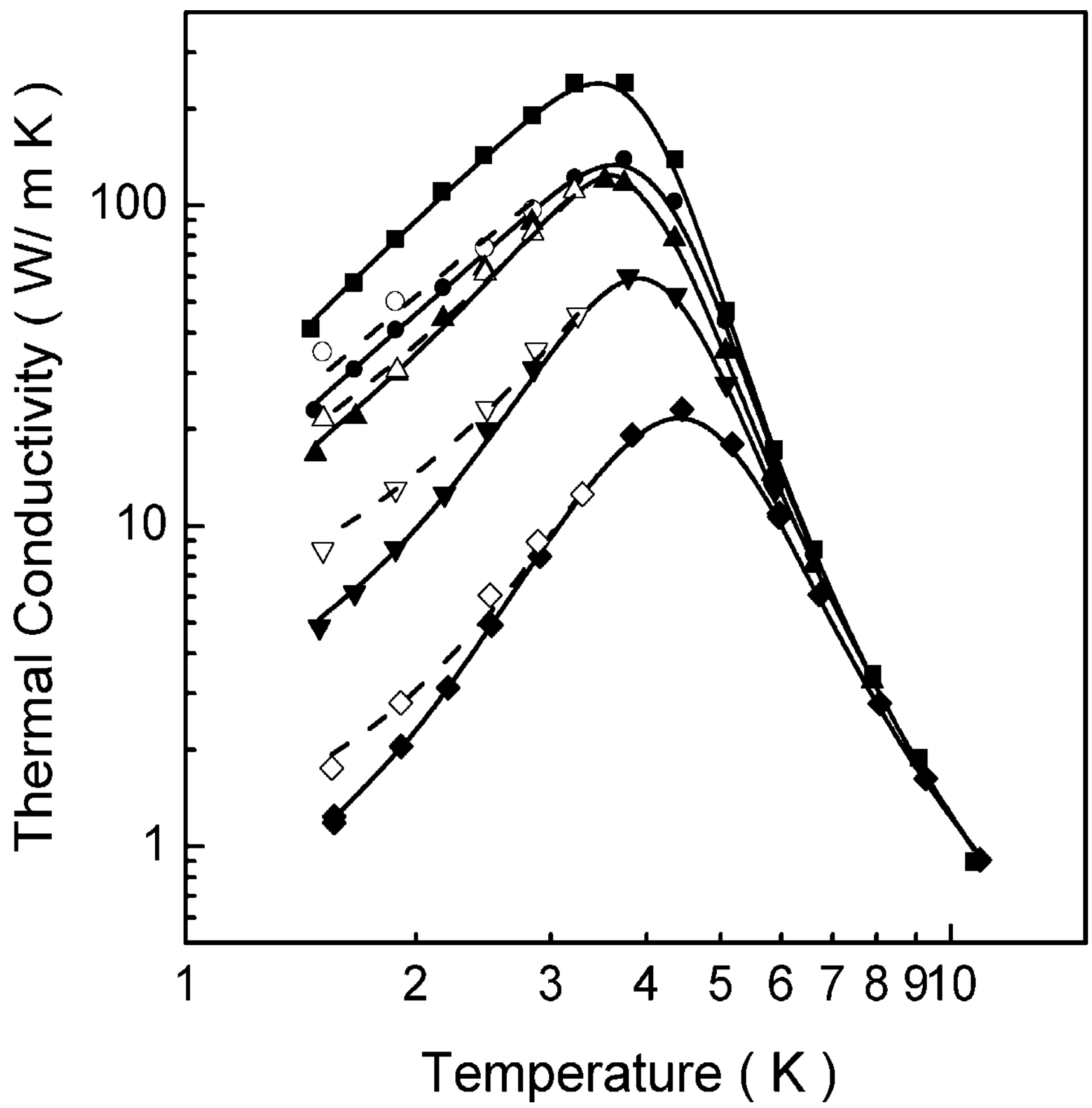

FIG. 3. Temperature dependences of the thermal conductivity of ortho-parahydrogen solutions[12] (•) $x$=0.21%, (•) $x$=0.56%, (A) $x$=0.96%, (V) $x$=2.4%, (•) $x$=4.4%. The open symbols correspond to a random distribution of the ortho subsystem.[18] Solid and dash lines – $K$-curves computed for equilibrium and random distributions.

Ref. 20 as

$$\tau_N^{-1} = 6.7\cdot 10^3\, x^2 T^5\ \mathrm{c}^{-1}.$$

Resistive scattering includes the following mechanisms:

phonon-phonon scattering ($U$ processes)

$$\tau_U^{-1}(x,T) = A_U x^2 T^3 e^{(-E/T)};$$

where $E$ is the threshold phonon energy at which Umklapp processes first appear;

boundary scattering

$$\tau_B^{-1} = L/s;$$

where $L$ is the characteristic mean-free path length for boundary scattering;

scattering by isolated orthomolecules[9]

$$\tau_s^{-1} = C\, n_s(xT + 3x^2T^2)$$

where the coefficient $C$ is an adjustable parameter in this model.

Klein's expression for scattering of phonons by a two-level system,[21] summed over each pair of levels in the rotational spectrum, was used to describe the scattering by a pair cluster with four nine-fold degenerate levels (see Fig. 1):

$$\tau_p^{-1} = \frac{2\pi\hbar^2 s^3 N n_p}{9k^2 V} \sum_i \frac{d_i S_i(T)}{\omega_{ri}^2} \frac{\gamma(0)\gamma(T)(xT/\omega_{ri})^4}{(1-(xT/\omega_{ri})^2)^2 + \gamma(T)^2 (xT/\omega_{ri})^6}$$

where $N$ is Avogadro number; $V$ is the molar volume; $n_s$ and $n_p$ are, respectively, the concentration of singles and pairs; $d_i$ and $S_i(T)$ are, respectively, the degeneracy and the population difference of each of the two levels included in a transition; $\omega_{ri}$ is the transition frequency; $\gamma$ (0) is an adjust-able parameter; $\gamma$ ($T$) is the total width of the two levels, which can vary substantially from level to level as a result of changes in the selection rules in the presence of an interaction of phonons of different symmetry.[21] In the present analysis the temperature dependence $\gamma$ ($T$) was chosen empirically to be the same for all levels and is represented as

$$\gamma(T) = \gamma(0)\, \mathrm{cth}^2 (\omega_{ri} / 8T).$$

The parameters of the $U$-processes for each crystal were chosen beforehand according to the high-temperature branch of the thermal-conductivity curve and were kept constant in the subsequent analysis. The results of the fit are shown in Fig. 3: the solid lines represent experimental quasiequilibrium curves and the broken lines represent the thermal conductivity with a high-temperature distribution of the ortho subsystem. When fitting the thermal conductivity of the crystal using a high-temperature distribution for the ortho subsystem, the concentrations of singles $n_s$, and pairs $n_p$ in the expressions (1) and (2) corresponded to a random distribution. To calculate $n_s$, and $n_p$ the expressions from Ref. 22 were used for the fit. Thus, three adjustable parameters were used to analyze the experimental curves of the thermal conductivity: *s/L,* *C,* and $\gamma$ (0). The values obtained for the adjustable parameters are presented in Table I.

TABLE I. Parameters of phonon scattering mechanisms, obtained by fitting the experimental data.

| $x$ | $s/L$, s$^{-1}$ | $C$, s$^{-1}$ | $\gamma$ (0) | $A_U$, s$^{-1}\cdot$K$^{-3}$ | $E$.K |
|---|---|---|---|---|---|
| 0.0021 | $1.175\cdot10^6$ | $4.225\cdot10^9$ | $3.820\cdot10^{-5}$ | $7.663\cdot10^7$ | 41.81 |
| 0.0050 | $1.987\cdot10^6$ | $3.676\cdot10^9$ | $3.820\cdot10^{-5}$ | $6.890\cdot10^7$ | 41.05 |
| 0.0050* | $5.004\cdot10^5$ | $7.8\cdot10^9$ | $3.820\cdot10^{-5}$ | $6.890\cdot10^7$ | 41.05 |
| 0.0096 | $4.076\cdot10^6$ | $1.002\cdot10^7$ | $3.820\cdot10^{-5}$ | $3.871\cdot10^7$ | 36.7 |
| 0.0096* | $3.0068\cdot10^6$ | $3.1429\cdot10^9$ | $3.820\cdot10^{-5}$ | $3.871\cdot10^7$ | 36.7 |
| 0.0210 | $1.594\cdot10^7$ | $1.030\cdot10^6$ | $3.820\cdot10^{-5}$ | $4.666\cdot10^7$ | 37.24 |
| 0.0210* | $8.101\cdot10^6$ | $3.385\cdot10^9$ | $3.820\cdot10^{-5}$ | $4.666\cdot10^7$ | 37.24 |
| 0.0440 | $7.763\cdot10^7$ | $1.0019\cdot10^7$ | $3.820\cdot10^{-5}$ | $5.261\cdot10^7$ | 38.26 |
| 0.0440* | $5.535\cdot10^7$ | $3.1965\cdot10^9$ | $3.820\cdot10^{-5}$ | $5.261\cdot10^7$ | 38.26 |

*Note:* The values of $x$ marked by an asterisk correspond to a random distri-bution of the ortho subsystem.

## 3. CONCLUSIONS

We shall not discuss the phonon relaxation mechanisms that give thermal conductivity proportional to $T^3$ and $T^2$. These mechanisms can be determined by the dislocation structure of the experimental crystals and their intensity could likewise depend on the content of the ortho modification. The frequency dependence of these mechanisms is close to that of phonon scattering by dislocations. The mechanisms due to structural defects have been discussed in an investigation of isotopic parahydrogen - orthodeuterium solutions.[4] The presence of orthomolecules makes hydrogen a more rigid crystal than pure parahydrogen;[13] this could substantially increase the dislocation density with temperature cycling, which was performed in the experiment to obtain a relaxation effect.[8]

Figure 4 shows the comparative contributions of the scattering mechanisms for a sample with a concentration of orthomolecules $x = 0.98\%$. The contribution of scattering by pairs of orthomolecules to the thermal conductivity with the number of pairs varying from random to equilibrium is shown in Fig. 3 (solid and broken lines). Resonance scattering by pairs makes a substantial contribution to phonon scattering. Resonance scattering between the second and third levels in the rotational spectrum, which are split by 0.83 K, makes the main contribution, which determines the behavior of the thermal conductivity in the experimental temperature range. It should be noted that the purpose of the present analysis was not to describe the experiment precisely taking account all details, for example, conversion, the contribution of triplets of orthomolecules to scattering, and so on.

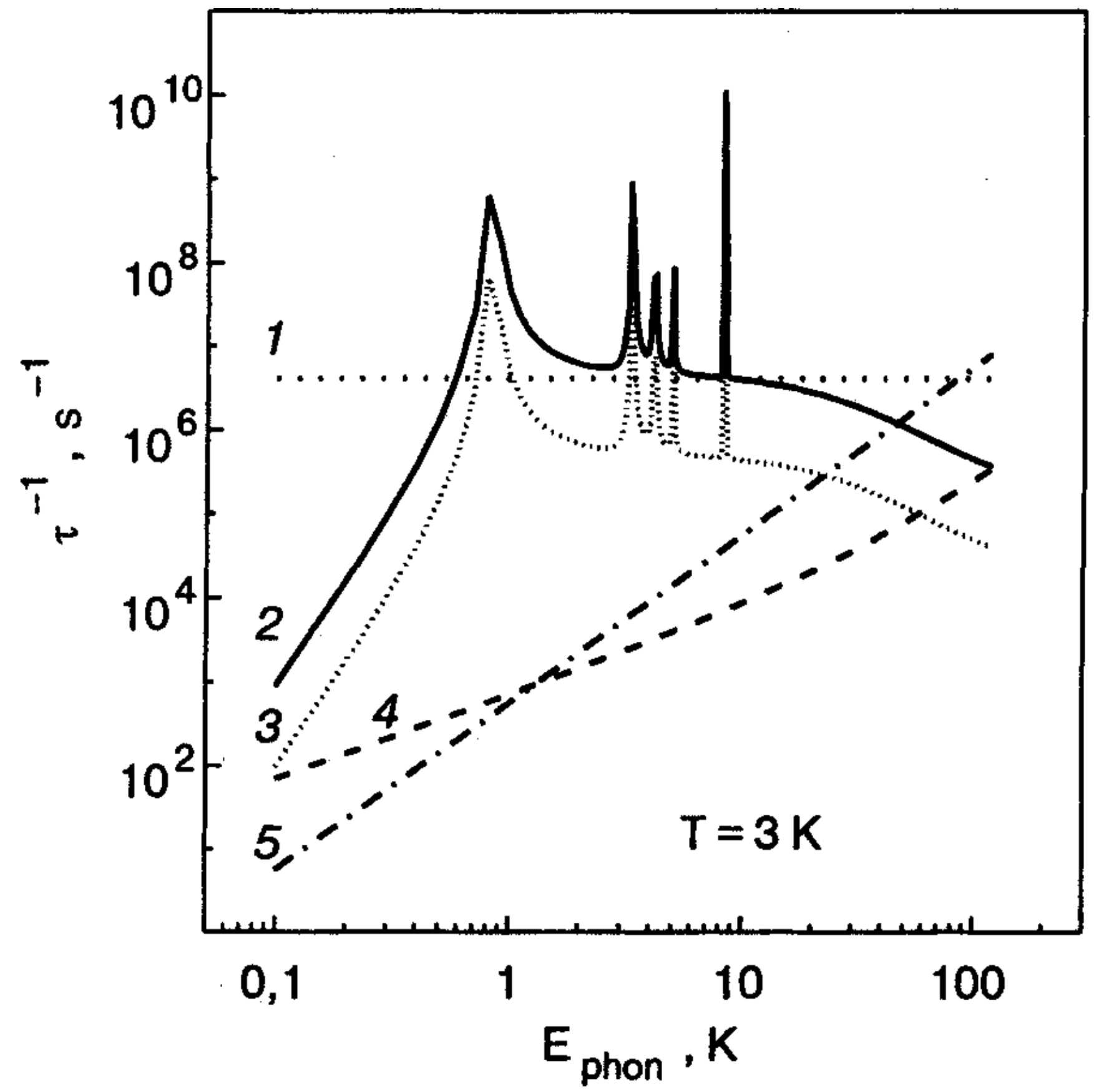

FIG. 4. Comparative contributions of phonon scattering mechanisms as a function of phonon energy for a sample with $x$= 0.98% with an equilibrium distribution. *(1)*—boundary scattering; (2)—scattering by pair clusters (equilibrium distribution); *(3)*—scattering by pair clusters (random distribution); *(4)*—scattering by singles; (5)—phonon-phonon scattering.

In summary, it has been shown in this work that the observed change due to orthomolecules in the thermal conductivity can be described on the basis of resonance scattering of phonons by pairs of orthomolecules. The present analysis is undeniably empirical and qualitative, but elaboration of the theory expounded in Ref. 14 could solve the problem of how the ortho impurity affects the thermal conductivity of parahydrogen.

E-mail: gorodilov@ilt.kharkov.ua